\documentclass{article}
\usepackage{amssymb}

\usepackage{fleqn}
\usepackage{psfig}
\usepackage{epsfig}
\usepackage{graphicx}
\usepackage{amsmath}

\input{tcilatex}
\def\beq{\begin{equation}}
\def\eeq{\end{equation}}
\def\bea{\begin{eqnarray}}
\def\eea{\end{eqnarray}}
\def\bq{\begin{quote}}
\def\eq{\end{quote}}

\def\gappeq{\mathrel{\rlap {\raise.5ex\hbox{$>$}}
{\lower.5ex\hbox{$\sim$}}}}
\def\lappeq{\mathrel{\rlap{\raise.5ex\hbox{$<$}}
{\lower.5ex\hbox{$\sim$}}}}
\def\Toprel#1\over#2{\mathrel{\mathop{#2}\limits^{#1}}}

\begin{document}

\pagestyle{empty} 
\begin{flushright}
{CERN-TH/2001-050}\\
hep-ph/0103002\\
\end{flushright}
\vspace*{15mm}

\begin{center}
\textbf{NEXT-TO-LEADING RESUMMED COEFFICIENT FUNCTION\\[2pt]
FOR THE SHAPE FUNCTION } \\[9pt]
\vspace{2cm} $~~$\\[0pt]
\textbf{U. Aglietti}$^{\ast )}$ \\[0pt]
\vspace{0.1cm} $~~$\\[0pt]
Theoretical Physics Division, CERN\\[1pt]
CH - 1211 Geneva 23 \\[3pt]
$~~$ \\[0pt]
\vspace*{1.5cm} \textbf{Abstract} \\[0pt]
\end{center}

We present a next-to-leading evaluation of the resummed coefficient function
for the shape function. The results confirm our previous leading-order
analysis, namely that the coefficient function is short-distance-dominated,
and allow relating the shape function computed with a non-perturbative
technique to the physical QCD distributions. \vspace*{5cm} \noindent 

\noindent $^{\ast )}$ On leave of absence from Dipartimento di Fisica,
Universita' di Roma I, Piazzale Aldo Moro 2, 00185 Roma, Italy. E-mail
address: ugo.aglietti@cern.ch. \vspace*{0.5cm}

\begin{flushleft} CERN-TH/2001-050 \\[0pt]
March 2001
\end{flushleft}
\vfill\eject

\setcounter{page}{1} \pagestyle{plain}

\section{Introduction}

In this note we present a next-to-leading evaluation of the coefficient
function of the shape function. The leading-order analysis was done in \cite
{eccomi}. The coefficient function allows relating the shape function ---
computed with a non-perturbative technique --- to physical distributions in
semi-inclusive heavy-flavour decays.

In general, we consider the process 
\begin{equation}
H_{h}\rightarrow X+\left( \mathsf{non-QCD\,partons}\right)
\label{generalizzo}
\end{equation}
in the hard limit 
\begin{equation}
Q\gg \Lambda  \label{hardlim}
\end{equation}
and in the semi-inclusive region\footnote{%
This region is also called threshold region, large-$x$ region,
radiation-inhibited region and Sudakov region.} 
\begin{equation}
M\ll Q,  \label{seminc}
\end{equation}
where $M$ is the invariant mass of $X,$ the hadronic final state; $H_{h}$ is
a hadron containing a heavy quark $h$ and the non-QCD partons can be a
lepton pair, a vector boson, a photon, etc. The quantity $Q$ is the hard
scale of the time-like process: $Q\equiv 2\mathcal{E}$, where $\mathcal{E}$
is the final hadronic energy and $\Lambda $ is the QCD scale. Well-known
examples of (\ref{generalizzo}) are: 
\begin{equation}
B\rightarrow X_{s}+\gamma  \label{rare}
\end{equation}
for a large photon energy and 
\begin{equation}
B\rightarrow X_{u}+l+\nu  \label{semi}
\end{equation}
for a small hadronic mass or a large electron energy\footnote{%
For the rare decay (\ref{rare}), since $Q=m_{B}\left(
1+m_{X_{s}}^{2}/m_{B}^{2}\right) ,$ one can actually set $Q=m_{B}.$}.

For the hadron at rest, i.e. with velocity $v=\left( 1;0,0,0\right) ,$ and
the jet $X$ flying along the minus direction ($-z$ axis), the shape function
is defined as \cite{nostri}--\cite{separati}: 
\begin{equation}
\varphi \left( k_{+}\right) \equiv \langle H_{h}\left( v\right)
|h_{v}^{\dagger }\,\delta \left( k_{+}-iD_{+}\right) h_{v}|H_{h}\left(
v\right) \rangle .  \label{shape}
\end{equation}
The latter is also called structure function of the heavy flavour and
represents the probability that the heavy quark in the hadron has momenta 
\begin{equation}
p_{h}=m_{H}v+k  \label{perme}
\end{equation}
with any transverse and minus component and with given plus component $%
k_{+}. $ The static field $h_{v}\left( x\right) $ is related to the Dirac
field of the heavy quark $h\left( x\right) $ by: 
\begin{equation}
h\left( x\right) =e^{-im_{H}v\cdot x}h_{v}\left( x\right) +O\left( \frac{%
\Lambda }{m_{H}}\right) .
\end{equation}
With the shape function, the decay (\ref{generalizzo}) is related to its
respective quark-level process 
\begin{equation}
h\rightarrow \widehat{X}+\left( \mathsf{non-QCD\,partons}\right) ,
\end{equation}
where the heavy quark has the momentum (\ref{perme}) with the distribution (%
\ref{shape}). Note that the state $\widehat{X}$, unlike $X,$ does not
contain the light valence quark(s) in $H.$ The invariant mass $m$ of $%
\widehat{X}$ is related to the plus virtuality of the heavy quark $k_{+}$\
by the relation: 
\begin{equation}
k_{+}=-\frac{m^{2}}{Q}.
\end{equation}
The shape function is a non-perturbative distribution --- analogous to
parton distribution functions --- and it describes the slice of the
semi-inclusive region in which 
\begin{equation*}
m^{2}\sim \Lambda \,Q.
\end{equation*}
In terms of the variables used in QCD and in the effective theory (ET), this
is: 
\begin{equation}
k_{+}\sim \Lambda ,\mathrm{\quad }\quad z\sim 1-\frac{\Lambda }{Q},\mathrm{%
\quad }\quad N\sim \frac{Q}{\Lambda },  \label{cinshape}
\end{equation}
where 
\begin{equation}
z\equiv 1-\frac{m^{2}}{Q^{2}},
\end{equation}
and $N$ is the moment index (see later).

The shape function describes the mass distribution \ of a jet coming from
the decay of a heavy flavour. It therefore generalizes to region (\ref
{cinshape}) the jet function $J\left( Q^{2},m^{2}\right) $ introduced in
ref. \cite{cattren}, representing the probability that a light parton
produced in a hard collision with scale $Q$ evolves inclusively into a jet
with mass $m.$ The main difference is that in our case the distribution
depends on the process as a whole, and not only on the fragmentation of the
final quark. The shape function depends on the initial hadron state; for
example, it is different for a $B$ meson and a $\Lambda _{b}$ hyperon.

The coefficient function $C$ is defined by the relation\footnote{%
QCD quantities are denoted by bold-face symbols, the related quantities in
the ET by normal symbols.} 
\begin{equation}
\mathbf{\varphi }\left( k_{+};Q\right) =\int dk_{+}^{\prime }\,C\left(
k_{+}-k_{+}^{\prime };Q,\mu \right) \,\varphi \left( k_{+}^{\prime };\mu
\right) ,
\end{equation}
where $\mu $ is the renormalization point or UV cut-off of the operator
entering the definition of the shape function (see eq.\thinspace (\ref{shape}%
)). The coefficient function is then obtained by evaluating in
next-to-leading order (NLO) the QCD effective form factor $\mathbf{\varphi }%
\left( k_{+};Q\right) $ and the shape function $\varphi \left( k_{+};\mu
\right) $ and taking their ratio (see later). Since $C$ is expected to be a
short-distance quantity, we compute the QCD distribution and the shape
function in perturbation theory (PT) for an on-shell heavy quark $(k=0).$
This expectation will be verified \textit{a posteriori. }

The coefficient function is different, even at the leading level, for the
shape function regulated in dimensional regularization (DR) and in lattice
regularization; it is short-distance-dominated in both cases \cite{myproc}.
The main difference is that, at one loop, $C$ contains a double logarithm of 
$k_{+}$ in DR, while it contains at most a single logarithm of $k_{+}$ in
lattice regularization. In view of the applications, we specialize ourselves
in the lattice case.

The paper is organized as follows. In sec.\thinspace 2 we present a compact
derivation of a resummed QCD form factor $\mathbf{\varphi }\left(
k_{+};Q\right) $ to NLO accuracy. In sec.\thinspace 3 we perform a similar
computation for the shape function $\varphi \left( k_{+};\mu \right) .$ In
sec.\thinspace 4 we evaluate the coefficient function and we discuss the
physical implications. Finally, in sec.\thinspace 5, we draw our conclusions.

\section{The QCD distribution}

We are interested in a general QCD distribution $\mathbf{f}\left( z\right) $
in the threshold region. For clarity's sake, let us consider a specific
case: the photon spectrum in the decay (\ref{rare}), 
\begin{equation}
\mathbf{f}\left( x\right) =\frac{1}{\Gamma _{B}}\frac{d\Gamma }{dx},
\end{equation}
where 
\begin{equation}
x\equiv \frac{2E_{\gamma }}{m_{b}}
\end{equation}
and $\Gamma _{B}$ is the Born width: 
\begin{equation}
\Gamma _{B}=\frac{\alpha }{\pi }\frac{G_{F}^{2}m_{b}^{3}
m_{b,\overline{MS}}^2(m_b) }{32\pi ^{3}}%
|V_{tb}V_{ts}^{\ast }|^{2}|C_{7}\left( \mu _{b}\right) |^{2}.
\end{equation}
The $\overline{MS}$ mass is related to the pole-mass by:
\begin{equation}
m_{b,\overline{MS}}(m_b)= \left(1-\frac{\alpha_S C_F }{\pi }\right) m_b.
\end{equation}
The distribution reads, to order $\alpha _{S}$ \cite{citare}: 
\begin{equation}
\mathbf{f}\left( x\right) =\overline{\mathbf{f}}\left( x\right) +\alpha
_{S}\,k\,\delta \left( 1-x\right) +\alpha _{S}\,d\left( x\right) .
\label{cominciamento}
\end{equation}
The function $\overline{\mathbf{f}}\left( x\right) $ factorizes
(perturbative) long-distance effects occurring in (\ref{rare}) and reads:%
\footnote{%
Overlined quantities denote subtracted quantities, containing only infrared
logarithms.} 
\begin{equation}
\overline{\mathbf{f}}\left( x\right) =\delta \left( 1-x\right) -A_{1}\alpha
_{S}\left( \frac{\log \left[ 1-x\right] }{1-x}\right) _{+}+B_{1}\alpha
_{S}\left( \frac{1}{1-x}\right) _{+},  \label{verificata}
\end{equation}
where 
\begin{equation}
A_{1}=\frac{C_{F}}{\pi },\qquad \qquad B_{1}=-\frac{7}{4}\frac{C_{F}}{\pi }
\end{equation}
and $C_{F}=\left( N_{c}^{2}-1\right) /\left( 2N_{c}\right) =4/3.$ The spike
term involves the constant 
\begin{equation}
k=-\frac{C_{F}}{\pi }\left( \log \frac{\mu _{b}}{m_{b}}+\frac{5}{4}+\frac{%
\pi ^{2}}{3}\right) ,
\end{equation}
where $\mu _{b}$ is the renormalization scale of the operator $\mathcal{O}%
_{7};$ this (unphysical) scale must be taken of $O\left( m_{b}\right) $ in
order to avoid large (ultraviolet) logarithms in the matrix elements. The
function $d\left( x\right) $ is the ``remainder'': 
\begin{equation}
d\left( x\right) =\frac{C_{F}}{4\pi }\left[ 7+x-2x^{2}-2\left( 1+x\right)
\log \left( 1-x\right) \right] .
\end{equation}
Plus-distributions are defined as usual as $P\left( x\right) _{+}\equiv
P\left( x\right) \,-\,\delta \left( 1-x\right) \int_{0}^{1}dy\,P\left(
y\right) .$ For the resummation, it is natural to write the function $%
\overline{\mathbf{f}}\left( x\right) $ in an ``unintegrated'' form as 
\begin{equation}
\overline{\mathbf{f}}\left( x\right) =\delta \left( 1-x\right)
+\int_{0}^{1}d\epsilon \int_{0}^{1}dt\left[ \frac{A_{1}\alpha _{S}}{\epsilon
t}+\frac{S_{1}\alpha _{S}}{\epsilon }+\frac{C_{1}\alpha _{S}}{t}\right] 
\left[ \delta \left( 1-x-\epsilon t\right) -\delta \left( 1-x\right) \right]
,  \label{bo}
\end{equation}
where we have defined the unitary energy and angular variables 
\begin{equation}
\epsilon \equiv \frac{E}{Q}\qquad \mathrm{and}\qquad t\equiv \frac{1-\cos
\theta }{2}.
\end{equation}
The quantity $E$ is two times the energy of the soft gluon, $E=2E_{g},$ and $%
\theta $ is the gluon emission angle. It proves to be convenient to separate
the soft from the collinear term, as we have done. By explicit evaluation,
we find 
\begin{equation}
S_{1}=-\frac{C_{F}}{\pi },\qquad \qquad C_{1}=-\frac{3}{4}\frac{C_{F}}{\pi }.
\end{equation}
Note that $B_{1}=S_{1}+C_{1},$ i.e. it is the sum of the coefficients of the
soft and the collinear terms. Expression (\ref{bo}) is symmetric for $%
\varepsilon \leftrightarrow t.$

Next-to-leading corrections are included following these prescriptions:

\begin{enumerate}
\item  Replacement of the bare coupling by the two-loop running coupling: 
\begin{equation}
\alpha _{s}\left( q^{2}\right) =\frac{1}{\beta _{0}\log q^{2}/\Lambda ^{2}}-%
\frac{\beta _{1}}{\beta _{0}^{3}}\,\frac{\log \log q^{2}/\Lambda ^{2}}{\log
^{2}q^{2}/\Lambda ^{2}},
\end{equation}
evalutated at the gluon transverse momentum squared \cite{ven}: 
\begin{equation}
\alpha _{S}\rightarrow \alpha _{S}\left( k_{\perp }^{2}\right) ,
\end{equation}
where 
\begin{equation}
k_{\perp }^{2}\simeq Q^{2}\epsilon ^{2}t.
\end{equation}
The first two coefficients of the $\beta $-function are: 
\begin{eqnarray}
\qquad \qquad \qquad \qquad \beta _{0} &=&\frac{11C_{A}-2n_{F}}{12\pi }=%
\frac{33-2n_{F}}{12\pi },  \notag \\
\beta _{1} &=&\frac{17C_{A}^{2}-5C_{A}n_{F}-3C_{F}n_{F}}{24\pi ^{2}}=\frac{%
153-19\,n_{F}}{24\pi ^{2}},
\end{eqnarray}
where $C_{A}=N_{c}=3$ and $n_{F}=3$ is the number of active quark flavours;

\item  Inclusion of the two-loop correction to the term $A_{1}\alpha _{S},$
so that 
\begin{equation}
A_{1}\alpha _{S}\rightarrow A_{1}\alpha _{S}+A_{2}\alpha _{S}^{2}.
\end{equation}
The explicit computation gives \cite{kodtren, cattren} 
\begin{equation}
A_{2}=\frac{1}{2}\frac{C_{F}}{\pi ^{2}}K
\end{equation}
where, in the $\overline{MS}$ \ scheme for the coupling constant, 
\begin{equation}
K=C_{A}\left( \frac{67}{18}-\frac{\pi ^{2}}{6}\right) -\frac{10}{9}%
n_{f}T_{R},
\end{equation}
with $T_{R}=1/2.$
\end{enumerate}

The ``effective'' one-gluon distribution\footnote{%
We call it ``effective'' because the insertion of the running coupling in
the time-like region and of the term proportional to $A_{2}$ already
includes some multiple-gluon-emission effects.} then reads: 
\begin{eqnarray}
\overline{\mathbf{f}}\left( x\right) &=&\delta \left( 1-x\right)
+\int_{0}^{1}d\epsilon \int_{0}^{1}dt\left[ \frac{A_{1}\alpha _{S}\left(
Q^{2}\epsilon ^{2}t\right) +A_{2}\,\alpha _{S}^{2}\left( Q^{2}\epsilon
^{2}t\right) }{\epsilon t}+\frac{S_{1}\,\alpha _{S}\left( Q^{2}\epsilon
^{2}t\right) }{\epsilon }+\frac{C_{1}\,\alpha _{S}\left( Q^{2}\epsilon
^{2}t\right) }{t}\right]  \notag \\
&&\qquad \qquad \qquad \qquad \qquad \left[ \delta \left( 1-x-\epsilon
t\right) -\delta \left( 1-x\right) \right] .
\end{eqnarray}
This integral is no longer symmetric for $\varepsilon \leftrightarrow t$
because of running coupling effects.

A consistent resummation to NLO accuracy (and beyond) is naturally performed
in moment space, so that we consider \cite{daqualcheparte}: 
\begin{eqnarray}
\qquad \qquad \qquad \qquad \qquad \qquad \overline{\mathbf{f}}_{N}
&=&\int_{0}^{1}dx\,\,x^{N-1}\,\overline{\mathbf{f}}\left( x\right)  \notag \\
&=&1\,+\,\Delta \overline{\mathbf{f}}_{N}.
\end{eqnarray}
With this definition of the Mellin transform, the total rate is given by the
first moment. Note that $\overline{\mathbf{f}}_{N=1}=1.$ The inverse
transform to the original distribution in $x$-space is usually done
numerically \cite{mangano}. The moments of the one-loop distribution $%
\overline{\mathbf{f}}\left( x\right) $ given in eq.\thinspace (\ref
{verificata}) read: 
\begin{equation}
\overline{\mathbf{f}}_{n}=1-\frac{A_{1}\alpha _{S}}{2}\log ^{2}n-B_{1}\alpha
_{S}\log n-\frac{A_{1}\pi ^{2}}{12}\alpha _{S}+O\left( \frac{1}{n}\right) .
\label{oneloop}
\end{equation}
We have defined $n\equiv N/N_{0},$ where $N_{0}\equiv e^{-\gamma
_{E}}=0.561459\ldots $ with $\gamma _{E}=0.577216\ldots $ the Euler constant 
\cite{cattren}. The single logarithm, the $\log n$ term, is positive and
tends to increase the rate, as is often the case \cite{lucavecchio}. Its
coefficient $B_{1}$ is rather big; it is over a factor of $2$ larger than in
DIS, for example, where $C_{1}$ is the same and $S_{1}=0.$ Simple
exponentiation of the effective one-gluon distribution takes place in $N$%
-space, so that 
\begin{equation}
\overline{\mathbf{f}}_{N}=e^{\Delta \overline{\mathbf{f}}_{N}}.
\end{equation}
The non-logarithmic integrations of the terms proportional to $S_{1}$ and $%
C_{1}$ can be explicitly done \cite{cattren}\ and one obtains: 
\begin{eqnarray}
\log \overline{\mathbf{f}}_{N} &=&\int_{0}^{1}dx\,\frac{x^{N-1}-1}{1-x}%
\left\{ \int_{Q^{2}(1-x)^{2}}^{Q^{2}\left( 1-x\right) }\frac{dq^{2}}{q^{2}}%
\left[ \,A_{1}\alpha _{S}\left( q^{2}\right) +\,A_{2}\alpha _{S}^{2}\left(
q^{2}\right) \right] +\right.  \notag \\
&&\left. \qquad \quad \quad \quad +S_{1}\alpha _{S}\left( Q^{2}\left(
1-x\right) ^{2}\right) +C_{1}\alpha _{S}\left( Q^{2}(1-x)\right) \right\}
+O\left( \alpha _{S}^{n}\log ^{n-1}N\right) ,\qquad
\end{eqnarray}
Using the large-$N$ approximation $x^{N}-1\simeq -\theta \left(
1-x-1/n\right) $ \cite{cattren}\footnote{%
The error in this approximation amounts to next-to-next-to-leading
logarithms and to terms suppressed by powers of $1/N.$}, the moments read: 
\begin{eqnarray}
\log \overline{\mathbf{f}}_{n} &=&-\frac{A_{1}}{2\beta _{0}}\left[ \log
s\log \log s-2\log \frac{s}{n}\log \log \frac{s}{n}+\log \frac{s}{n^{2}}\log
\log \frac{s}{n^{2}}\right] +  \notag \\
&&+\frac{\beta _{0}A_{2}-\beta _{1}A_{1}}{2\beta _{0}^{3}}\left[ \log \log
s-2\log \log \frac{s}{n}+\log \log \frac{s}{n^{2}}\right] +  \notag \\
&&-\frac{\beta _{1}A_{1}}{4\beta _{0}^{3}}\left[ \log ^{2}\log s-2\log
^{2}\log \frac{s}{n}+\log ^{2}\log \frac{s}{n^{2}}\right] +  \notag \\
&&-\frac{S_{1}}{2\beta _{0}}\left[ \log \log s-\log \log \frac{s}{n^{2}}%
\right] -\frac{C_{1}}{\beta _{0}}\left[ \log \log s-\log \log \frac{s}{n}%
\right] ,\quad  \label{baba}
\end{eqnarray}
where $s$ is the square of the hard scale in unit of the QCD scale, 
\begin{equation}
s\equiv \frac{Q^{2}}{\Lambda ^{2}}.
\end{equation}
Note that $\overline{\mathbf{f}}_{n}=1$ for $n=1.$ Substituting $\log s$ in
eq.\thinspace (\ref{baba}) by its expression in terms of the two-loop
coupling, 
\begin{equation}
\log s=\frac{1}{\beta _{0}\alpha _{S}}+\frac{\beta _{1}}{\beta _{0}^{2}}\log
\left( \beta _{0}\alpha _{S}\right) ,
\end{equation}
one obtains the exponent (\ref{baba}) as a series of functions \cite
{cattren2}: 
\begin{equation}
\overline{\mathbf{f}}_{n}=\exp \left[ L\,\,g_{1}\left( \beta _{0}\alpha
_{S}L\right) +g_{2}\left( \beta _{0}\alpha _{S}L\right) +\alpha
_{S}\,g_{3}\left( \beta _{0}\alpha _{S}L\right) +\cdots \right] ,
\end{equation}
where the leading and next-to-leading functions are respectively \cite
{civuole}: 
\begin{eqnarray}
g_{1}\left( w\right) &=&-\frac{A_{1}}{2\beta _{0}}\,\frac{1}{w}\left[ \left(
1-2w\right) \log \left( 1-2w\right) -2\left( 1-w\right) \log \left(
1-w\right) \right] ;  \notag \\
\,g_{2}\left( w\right) &=&\frac{\beta _{0}A_{2}-\beta _{1}A_{1}}{2\beta
_{0}^{3}}\left[ \log (1-2w)-2\log (1-w)\right] -\frac{\beta _{1}A_{1}}{%
4\beta _{0}^{3}}\left[ \log ^{2}(1-2w)-2\log ^{2}(1-w)\right] +  \notag \\
&&+\frac{S_{1}}{2\beta _{0}}\log (1-2w)+\frac{C_{1}}{\beta _{0}}\log (1-w).
\end{eqnarray}
We have defined $\alpha _{S}\equiv \alpha _{S}\left( Q^{2}\right) $ and $%
L\equiv \log n.$ The two-loop terms coming from the expansion of the
resummed distribution read: 
\begin{equation}
\log \overline{\mathbf{f}}_{n}^{\,\,2loop}=-\frac{1}{2}A_{1}\beta
_{0}\,\alpha _{S}^{2}\,L^{3}-\left( \frac{1}{2}A_{2}+\beta _{0}S_{1}+\frac{1%
}{2}\beta _{0}C_{1}\right) \alpha _{S}^{2}\,L^{2}.  \label{twoloop}
\end{equation}
The effective form factor is real --- and therefore sensible --- only for 
\begin{equation}
n\,<\,n_{crit}\,=\,\exp \left[ \frac{1}{2\beta _{0}\alpha _{S}\left(
Q^{2}\right) }\right] \,\sim \,\frac{Q}{\Lambda }.
\end{equation}
This restriction is induced by the soft terms --- i.e. by those proportional
to $A_{1},S_{1}$ and $A_{2}.$ The reliability of the resummed result is then
limited to 
\begin{equation}
n\ll \frac{Q}{\Lambda }\qquad \Leftrightarrow \qquad m^{2}\gg \Lambda \,Q.
\label{limitati}
\end{equation}
The condition is the same as that in the leading order analysis. On the
other hand, the hard collinear contribution --- the term proportional to $%
C_{1}$ --- remains well-defined up to much larger values of \ $n:$%
\begin{equation}
n\ll \,\,n_{crit}^{2}\,\sim \,\frac{Q^{2}}{\Lambda ^{2}}\qquad
\Leftrightarrow \qquad m^{2}\gg \Lambda ^{2}.
\end{equation}
The collinear factor is therefore well-defined in the kinematic region (\ref
{cinshape}) described by the shape function. It becomes singular only in the
resonance or exclusive region $m^{2}\sim \Lambda ^{2}$, which is well beyond
the reach of the shape function.

\psfig{bbllx=90pt, bblly=335pt, bburx=650pt, bbury=740pt,
file=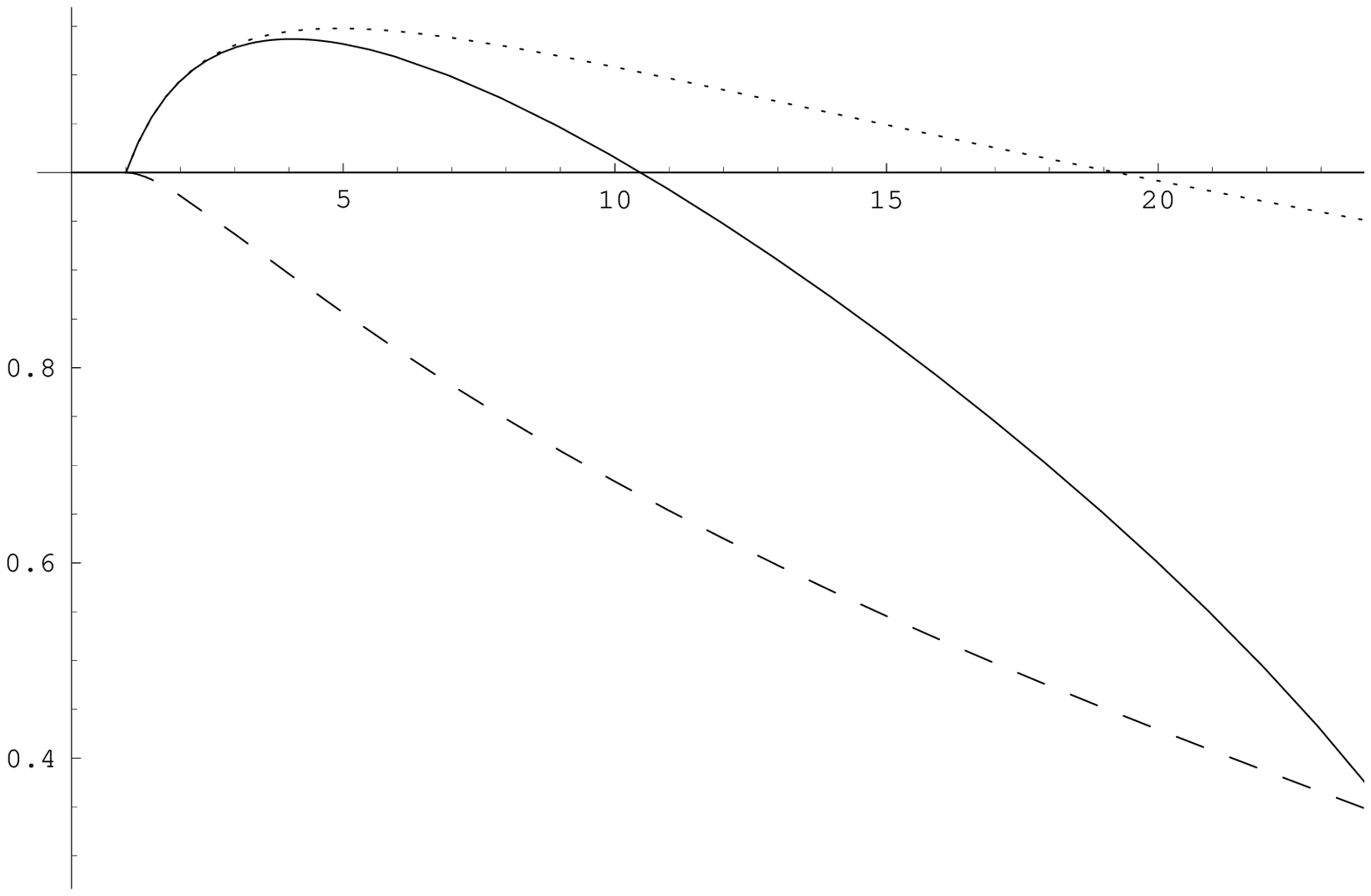, height=9cm, width=13cm} \vspace{-1.5cm}

\noindent Figure~1: Plot of the QCD form factor $\overline{\mathbf{f}}_{n}$
as a function of $n$ (first 24 moments) for the values of the parameters
discussed in the text. Solid line: NLO distribution; dashed line: LO
distribution; dotted line: expansion to order $\alpha _{S}^{2}$ of the
exponent in the NLO distribution.

\noindent The LO and NLO form factors are plotted in fig.\thinspace 1 for $%
Q=m_{B}\simeq 5.2$ GeV, for which $\alpha _{S}\simeq 0.21$ and $%
n_{crit}\,\simeq \,28.$ The NLO curve lies above the LO one and it exceeds
one in the small-$n$ region because of \ the single logarithm, as already
discussed. The two-loop curve lies above the NLO one because the leading
terms in higher orders are all negative (cf.\thinspace eqs.\thinspace (\ref
{oneloop}) and (\ref{twoloop})).

The resummed distribution is usually written as \cite{thrust}: 
\begin{equation}
\mathbf{f}_{N}\left( \alpha _{S}\right) \,=\,K\left( \alpha _{S}\right) \,%
\overline{\mathbf{f}}_{N}\left( \alpha _{S}\right) \,+\,D_{N}\left( \alpha
_{S}\right) \,,  \label{copio}
\end{equation}
where $K$ is a function with a power series expansion in $\alpha _{S}:$ 
\begin{equation}
K\left( \alpha _{S}\right) =1+k\,\alpha _{S}\,+k\,^{\prime }\alpha
_{S}^{2}\,+\cdots
\end{equation}
and $D_{N}$ is a remainder function: 
\begin{equation}
\quad D_{N}\left( \alpha _{S}\right) =\,d\,_{N}\,\alpha _{S}+d\,_{N}^{\prime
}\,\alpha _{S}^{2}+\cdots ,
\end{equation}
starting at order $\alpha _{S}$ and vanishing in the large-$N$ limit, 
\begin{equation}
D_{N}\sim \frac{1}{N}\qquad \mathrm{for}\qquad N\rightarrow \infty .
\end{equation}
To NLO accuracy, the contributions of order $\alpha _{S}$ to the function $K$
(and $D_{N})$\ are needed. To the same accuracy, in place of (\ref{copio}),
one can write $\mathbf{f}_{N}=(1+\alpha _{S}\,k+\alpha _{S}\,d\,_{N})\,%
\overline{\mathbf{f}}_{N}\,.$

An expression analogous to (\ref{copio}) also holds for the distributions in
the semileptonic decay (\ref{semi}) with a fixed hadronic energy, i.e. to
differential distributions in the hadronic energy. One simply has to make
the replacement 
\begin{equation}
x\rightarrow z
\end{equation}
in eq.\thinspace (\ref{copio}) and to account for a dependence of $K$ and $%
D_{N}$ on $Q/m_{B}.$ The function $\ \overline{\mathbf{f}}$ is the same in
the semileptonic and in the rare decay: 
\begin{equation}
\ \overline{\mathbf{f}}\ \left( x\right) =\overline{\mathbf{f}}\left(
z\right) .
\end{equation}
The general variable for the threshold region of heavy flavour decays is $z,$
even though, for the rare decay (\ref{rare}), it is more convenient to use $%
x.$ In the semi-inclusive region these two variables concide, because 
\begin{equation}
1-z=\frac{1-x}{\left( 2-x\right) ^{2}}=1-x+O\left( \left( 1-x\right)
^{2}\right) .
\end{equation}

\section{The shape function}

Let us now consider the quantity in the effective theory related to $\mathbf{%
f}\left( z\right) $, namely the shape function $\varphi \left( k_{+}\right)
. $ The latter has the analogous decomposition: 
\begin{equation}
\varphi \left( k_{+}\right) =\overline{\varphi }\left( k_{+}\right) +\alpha
_{S}\,h\,\delta \left( k_{+}\right) +\alpha _{S}\,p\left( k_{+}\right) .
\label{cheneso}
\end{equation}
The function $\overline{\varphi }\left( k_{+}\right) $ factorizes
long-distance effects and reads: 
\begin{eqnarray}
\overline{\varphi }\left( k_{+}\right) &=&\delta \left( k_{+}\right)
-A_{1}\alpha _{S}\,\frac{\theta \left( 0;k_{+};-\mu \right) }{-k_{+}}\log 
\frac{-k_{+}}{\mu }+A_{1}\alpha _{S}\,\delta \left( k_{+}\right) \int_{-\mu
}^{0}\frac{dl_{+}}{-l_{+}}\log \frac{-l_{+}}{\mu }+  \notag \\
&&\qquad \,\,\,+S_{1}\alpha _{S}\,\frac{\theta \left( 0;k_{+};-\mu \right) }{%
-k_{+}}-S_{1}\alpha _{S}\,\delta \left( k_{+}\right) \int_{-\mu }^{0}\frac{%
dl_{+}}{-l_{+}},
\end{eqnarray}
where $\mu \equiv 2\Lambda _{S}$ and we have defined $\theta \left(
a_{1};a_{2};\cdots a_{n}\right) \equiv \theta \left( a_{1}-a_{2}\right)
\theta \left( a_{2}-a_{3}\right) \cdots \theta \left( a_{n-1}-a_{n}\right) .$
The regularization used \cite{giuliame} imposes a cut-off on the spatial
loop momenta and not on the energies, 
\begin{equation}
|\overrightarrow{l}|<\Lambda _{S},\qquad -\infty <l_{0}<+\infty ,
\label{simpreg}
\end{equation}
and it is qualitatively similar to lattice regularization\footnote{%
By this we mean that the double logarithm at one loop is the same in the two
regularizations.}. Proceeding in a similar way, one can write: 
\begin{equation}
\overline{\varphi }\left( k_{+}\right) =\delta \left( k_{+}\right)
+\int_{0}^{\mu }\frac{dE}{E}\int_{0}^{1}dt\left[ \frac{A_{1}\alpha
_{S}\left( E^{2}t\right) +A_{2}\alpha _{S}^{2}\left( E^{2}t\right) }{t}%
+S_{1}\alpha _{S}\left( E^{2}t\right) \right] \left[ \delta \left(
k_{+}+Et\right) -\delta \left( k_{+}\right) \right] .  \label{ancoralei}
\end{equation}
A few comments are in order. The hard scale $Q$ does not appear in $%
\overline{\varphi }\left( k_{+}\right) $, as it should. The term
proportional to $C_{1}$ is absent because the shape function retains only
the leading terms in the soft limit $E\rightarrow 0,$ which are proportional
to $1/E.$ With the same definition of the coupling constant in QCD and in
the ET, the constant $A_{2}$ is the same in the two theories \cite{nuovaref,
nuovanuova}\footnote{%
A difference would imply a breakdown of factorization.}. Since $%
k_{+}=-Q\left( 1-z\right) ,$ the shape function in the ``QCD variable'' $z$
reads: 
\begin{equation}
\overline{f}\left( z\right) =\delta \left( 1-z\right) +\int_{0}^{1/r}\frac{%
d\epsilon }{\epsilon }\int_{0}^{1}dt\left[ \frac{A_{1}\alpha _{S}\left(
Q^{2}\epsilon ^{2}t\right) +A_{2}\,\alpha _{S}^{2}\left( Q^{2}\epsilon
^{2}t\right) }{t}+S_{1}\alpha _{S}\left( Q^{2}\epsilon ^{2}t\right) \right] 
\left[ \delta \left( 1-z-\epsilon t\right) -\delta \left( 1-z\right) \right]
,
\end{equation}
where we have defined an adimensional function as 
\begin{equation}
\overline{f}\left( z\right) \,\equiv \,Q\,\,\overline{\varphi }\left(
k_{+}\right) .
\end{equation}
The quantity $r$ is the ratio of the hard scale to the UV cut-off of the
effective theory, 
\begin{equation}
r\equiv \frac{Q}{\mu }>1,
\end{equation}
because the shape function is defined in a low-energy effective field
theory. To avoid substantial finite cut-off effects, one has also to assume 
\begin{equation}
\mu \gg \Lambda .  \label{ovvia}
\end{equation}
Taking moments and exponentiating the one-loop distribution as in the QCD
case, one obtains: 
\begin{equation}
\overline{f}_{N}=e^{\Delta \overline{f}_{N}}.
\end{equation}

\psfig{bbllx=90pt, bblly=392pt, bburx=610pt, bbury=740pt,
file=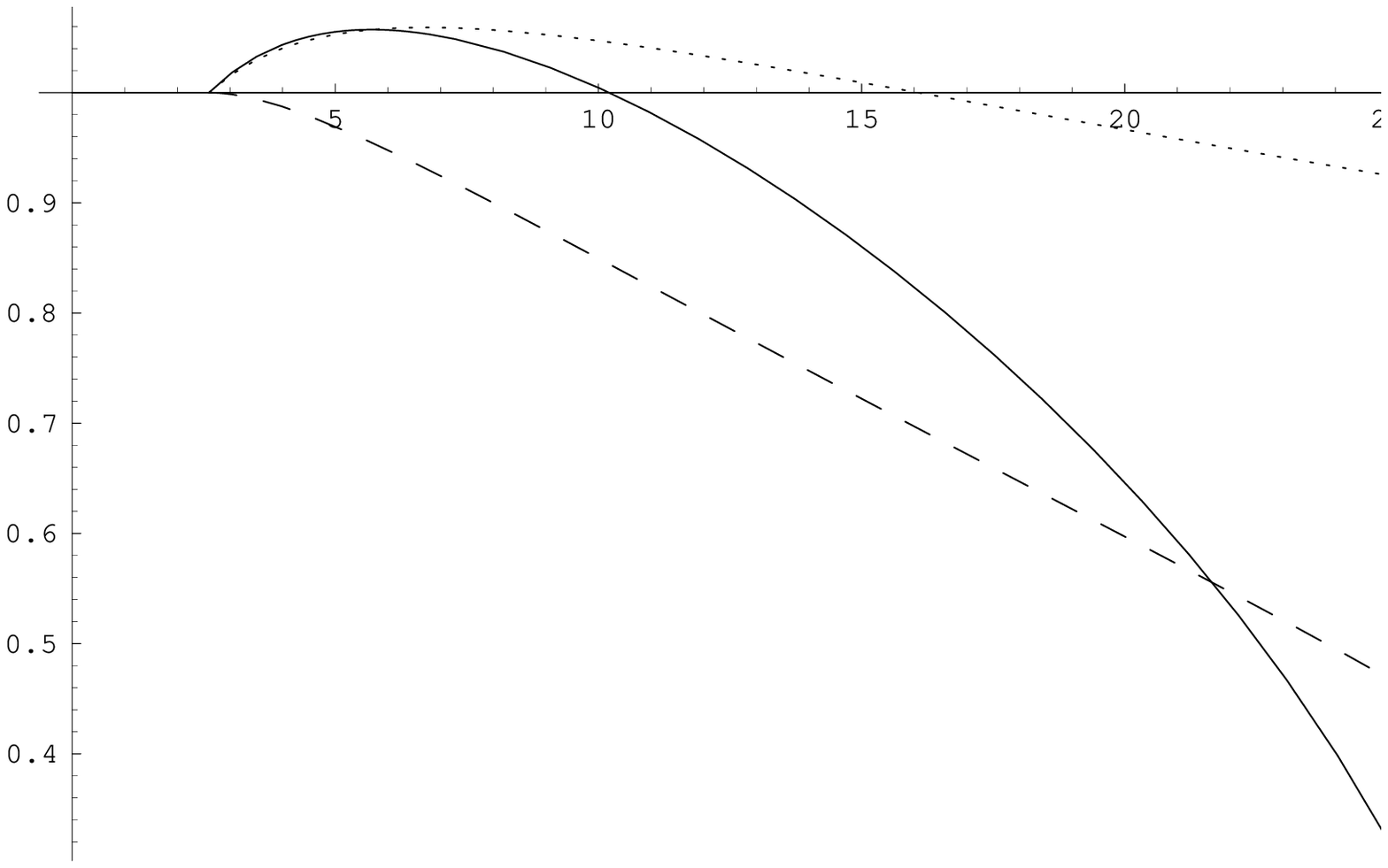, height=8cm, width=13cm}

\vspace{-0.5cm} \noindent Figure~2: Plot of the shape function $\overline{f}%
_{n}$ (first 25 moments). Solid line: NLO distribution; dashed line: LO
distribution; dotted line: expansion to order $\alpha _{S}^{2}$ of the
exponent in the NLO distribution.

\noindent Making the same approximations as in the QCD case, the shape
function is written: 
\begin{equation}
\log \,\overline{f}_{n}=-\theta \left( n-r\right) \int_{1/n}^{1/r}\frac{dy}{y%
}\left\{ \int_{y}^{1/r}\frac{d\epsilon }{\epsilon }\left[ A_{1}\alpha
_{S}\left( Q^{2}\epsilon y\right) +A_{2}\,\alpha _{S}^{2}\left(
Q^{2}\epsilon y\right) \right] +S_{1}\alpha _{S}\left( Q^{2}y^{2}\right)
\right\} .
\end{equation}
A straightforward integration then gives: 
\begin{eqnarray}
\log \,\overline{f}_{n} &=&\theta \left( n-r\right) \left\{ -\frac{A_{1}}{%
2\beta _{0}}\left[ \log \frac{s}{n^{2}}\log \log \frac{s}{n^{2}}-2\log \frac{%
s}{r\,n}\log \log \frac{s}{r\,n}+\log \frac{s}{r^{2}}\log \log \frac{s}{r^{2}%
}\right] +\right.  \notag \\
&&\qquad \qquad +\frac{A_{2}\beta _{0}-A_{1}\beta _{1}}{2\beta _{0}^{3}}%
\left[ \log \log \frac{s}{n^{2}}-2\log \log \frac{s}{r\,n}+\log \log \frac{s%
}{r^{2}}\right] +  \notag \\
&&\left. -\frac{A_{1}\beta _{1}}{4\beta _{0}^{3}}\left[ \log ^{2}\log \frac{s%
}{n^{2}}-2\log ^{2}\log \frac{s}{r\,n}+\log ^{2}\log \frac{s}{r^{2}}\right] +%
\frac{S_{1}}{2\beta _{0}}\left[ \log \log \frac{s}{n^{2}}-\log \log \frac{s}{%
r^{2}}\right] \right\} .
\end{eqnarray}

\psfig{bbllx=90pt, bblly=300pt, bburx=610pt, bbury=740pt,
file=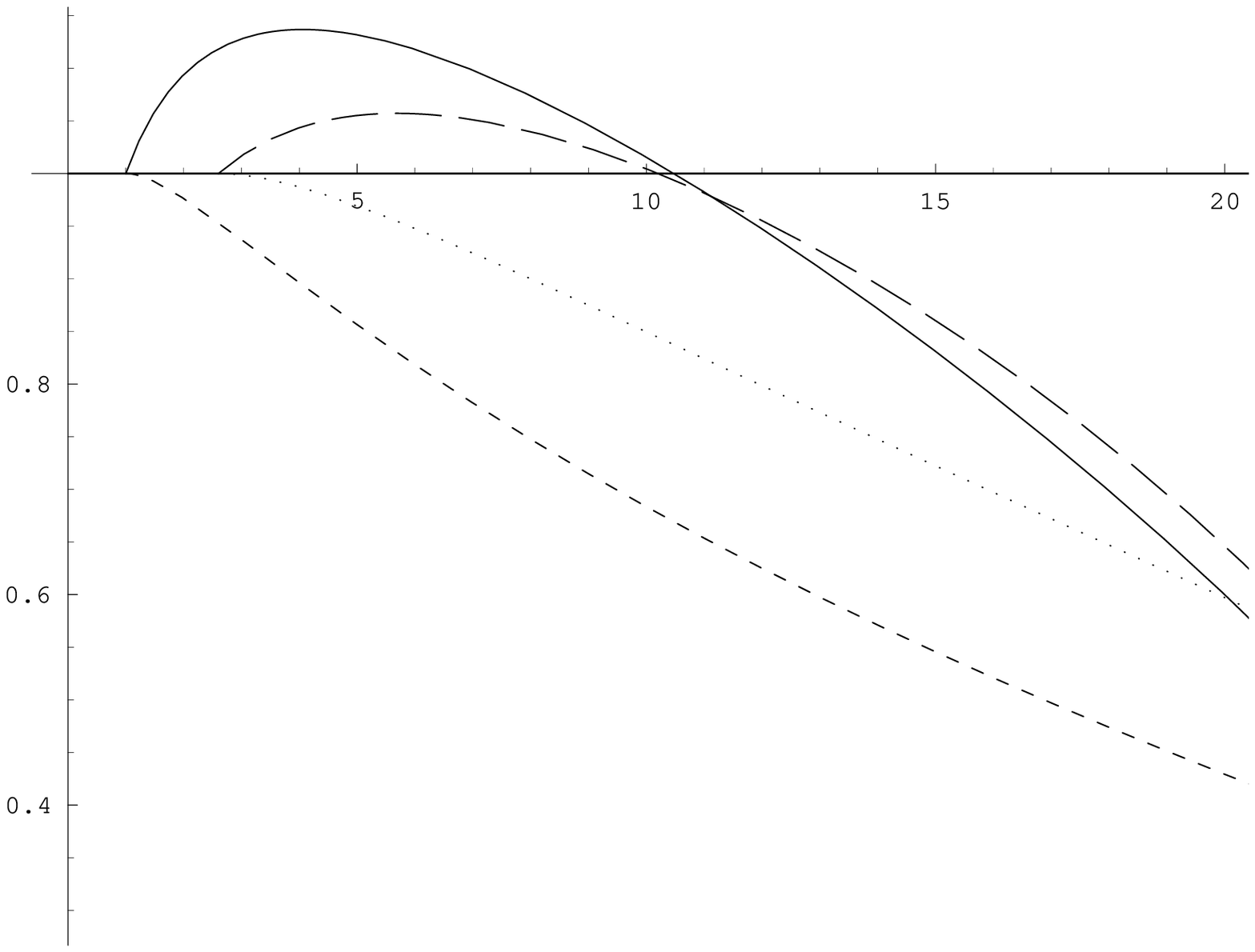, height=9cm, width=13cm} \vspace{-1 cm}

\noindent Figure~3: Comparison of QCD form factor with the shape function.
Solid line: NLO QCD distribution; big dashed line: NLO shape function; small
dashed line: LO QCD distribution; dotted line: LO shape function.

\noindent The main point is that the terms containing $n^{2}$ are
independent of $r:$ the infrared behaviour is the same as in QCD, as
expected on physical grounds. Substituting $\log s$ by its expression in
terms of the two-loop coupling, the result reads: 
\begin{equation}
\log \,\overline{f}_{n}\,=\,L\,\,g_{1}^{ET}\left( \beta _{0}\alpha
_{S}L;\,\beta _{0}\alpha _{S}R\right) \,+\,g_{2}^{ET}\left( \beta _{0}\alpha
_{S}L;\,\beta _{0}\alpha _{S}R\right) \,+\,\cdots  \notag
\end{equation}
where: 
\begin{eqnarray}
g_{1}^{ET}\left( w;\,\tau \right) &=&-\frac{A_{1}}{2\beta _{0}w}\left[
\left( 1-2w\right) \log \left( 1-2w\right) +\left( 1-2\tau \right) \log
\left( 1-2\tau \right) -2\left( 1-w-\tau \right) \log \left( 1-w-\tau
\right) \right] ,  \notag \\
g_{2}^{ET}\left( w;\,\tau \right) &=&\,\,\,\,\,\,\frac{A_{2}\beta
_{0}-A_{1}\beta _{1}}{2\beta _{0}^{3}}\left[ \log (1-2w)+\log \left( 1-2\tau
\right) -2\log \left( 1-w-\tau )\right) \right] +  \notag \\
&&-\frac{A_{1}\beta _{1}}{4\beta _{0}^{3}}\left[ \log ^{2}(1-2w)+\log
^{2}(1-2\tau )-2\log ^{2}\left( 1-w-\tau \right) \right] +  \notag \\
&&+\frac{S_{1}}{2\beta _{0}}\left[ \log (1-2w)-\log (1-2\tau )\right] ,
\end{eqnarray}
and 
\begin{equation}
R\equiv \log r>0,
\end{equation}
$\,$and the over-all factor $\theta \left( L-R\right) $ has been omitted.
Note that $\overline{f}_{n}$ is continuous at $n=r,$ but it has a cusp
singularity in this point. The range of $n$ is restricted by eq.\thinspace (%
\ref{limitati}): the same limitation of the QCD distribution occurs. The QCD
form factor $\overline{\mathbf{f}}_{n},$ except for the $C_{1}$ term, is
obtained by taking $\tau =0$ in the above expression, i.e. letting the gluon
energy to reach the hard scale. The shape function is plotted in
fig.\thinspace 2 for the same values of the parameters of the QCD
distribution and for $\mu =2\,$GeV. In fig.\thinspace 3 we compare the QCD
form factor with the shape function. In NLO, the shape-function curve is
below the QCD one for small $n$ because it has no $C_{1}$ term. NLO
distributions are very close to each other for large $n,$ but this seems
accidental; indeed, this phenomenon does not occur for the LO distributions.
In general, the shape-function curves look somehow shifted with respect to
the corresponding QCD ones.

The moments of the shape function can be decomposed in a form similar to the
QCD distribution: 
\begin{equation}
f_{N}\,\left( \alpha _{S}\right) =\,H\left( \alpha _{S}\right) \,\overline{f}%
_{N}\,\left( \alpha _{S}\right) +\,P_{N}\left( \alpha _{S}\right) \,,
\label{solita2}
\end{equation}
where the functions $\,H\left( \alpha _{S}\right) \,$\ and $P_{N}\left(
\alpha _{S}\right) $ have a power series expansion in $\alpha _{S}:$ 
\begin{eqnarray}
\qquad \qquad \qquad \qquad \qquad \qquad \qquad H\left( \alpha _{S}\right)
\,\, &=&1+h\,\alpha _{S}+h^{\prime }\alpha _{S}^{2}+\cdots  \notag \\
P_{N}\left( \alpha _{S}\right) &=&p_{N}\,\alpha _{S}+p_{N}^{\prime }\,\alpha
_{S}+\cdots .
\end{eqnarray}
To NLO accuracy, an expression alternative to (\ref{solita2}) including
higher-twist effects is $\ f_{N}=\left( 1+\alpha _{S}\,h\,+\alpha
_{S}\,p_{N}\right) \,\overline{f}_{N}.$

\section{The coefficient function}

We introduce the coefficient function $C_{N}$ relating the shape function to
the QCD form factor, as 
\begin{equation}
\mathbf{f}_{N}=\,C_{N}\,\,f_{N}.
\end{equation}
Inserting the expressions previously obtained and neglecting the remainder
functions $D_{N}$ and $P_{N}$, the coefficient function reads: 
\begin{equation}
C_{N}\,\left( \alpha _{S}\right) =\,G\left( \alpha _{S}\right) \,\Sigma
_{N}\left( \alpha _{S}\right) ,
\end{equation}
where 
\begin{equation}
G\left( \alpha _{S}\right) =\frac{K\left( \alpha _{S}\right) }{H\left(
\alpha _{S}\right) }=1+\,g\,\alpha _{S}+\cdots
\end{equation}
and 
\begin{equation}
\Sigma _{N}\left( \alpha _{S}\right) =\frac{\overline{\mathbf{f}}_{N}\left(
\alpha _{S}\right) }{\overline{\,f}_{N}\left( \alpha _{S}\right) }
\end{equation}
with $g=k-h.$ To NLO accuracy, the replacement $G\left( \alpha _{S}\right)
\rightarrow 1+\alpha _{S}\,g\,+\alpha _{S}\,\left( d_{N}-p_{N}\right) $ is
allowed, including also the remainder terms. The QCD distribution and the
shape function are related by a convolution in momentum space: 
\begin{equation}
\mathbf{f}\left( z\right) =\int_{0}^{1}\int_{0}^{1}dz^{\prime }dz^{\prime
\prime }\delta \left( z-z^{\prime }z^{\prime \prime }\right) \,C\left(
z^{\prime }\right) \,f\left( z^{\prime \prime }\right) .  \label{fondamento2}
\end{equation}
The function $\Sigma _{n}$ \ reads: 
\begin{equation}
\log \Sigma _{n}\,=\,\theta \left( r-n\right) \log \overline{\mathbf{f}}%
_{n}\,+\,\theta \left( n-r\right) \left[ \log \overline{\mathbf{f}}_{n}-\log
\,\overline{f}_{n}\right] .
\end{equation}
\newline
The explicit expression for $n>r$ is: 
\begin{equation}
\log \Sigma _{n}=\,L\,\,g_{1}^{CF}\left( \beta _{0}\alpha _{S}L;\,\beta
_{0}\alpha _{S}R\right) \,+\,g_{2}^{CF}\left( \beta _{0}\alpha _{S}L;\,\beta
_{0}\alpha _{S}R\right) \,+\,\cdots ,
\end{equation}
where 
\begin{eqnarray}
g_{1}^{CF}\left( w;\,\tau \right) &=&-\frac{A_{1}}{2\beta _{0}w}\left[
2\left( 1-w-\tau \right) \log \left( 1-w-\tau \right) -2\left( 1-w\right)
\log \left( 1-w\right) -\left( 1-2\tau \right) \log \left( 1-2\tau \right) %
\right] ;  \notag \\
g_{2}^{CF}\left( w;\,\tau \right) &=&\frac{A_{2}\beta _{0}-A_{1}\beta _{1}}{%
2\beta _{0}^{3}}\left[ 2\log \left( 1-w-\tau \right) -2\log (1-w)-\log
\left( 1-2\tau \right) \right] +  \notag \\
&&-\frac{A_{1}\beta _{1}}{4\beta _{0}^{3}}\left[ 2\log ^{2}\left( 1-w-\tau
\right) -2\log ^{2}(1-w)-\log ^{2}(1-2\tau )\right] +  \notag \\
&&+\frac{S_{1}}{2\beta _{0}}\log (1-2\tau )+\frac{C_{1}}{\beta _{0}}\log
(1-w).  \label{bella}
\end{eqnarray}

\psfig{bbllx=100pt, bblly=392pt, bburx=600pt, bbury=740pt,
file=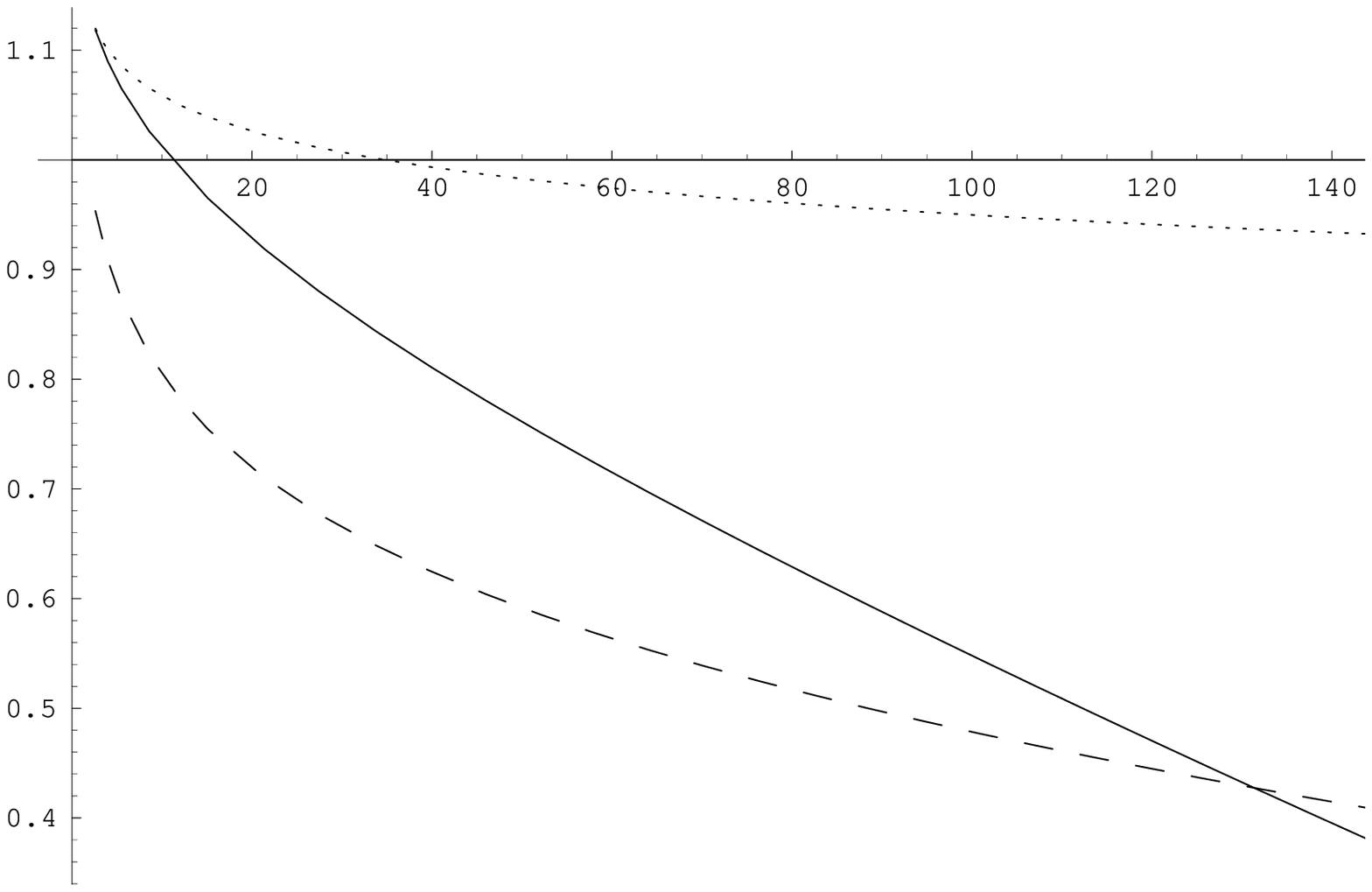, height=8cm, width=13cm }

\noindent Figure~4: Plot of the coefficient function $\Sigma _{n}$ for the
choice of the parameters discussed previously (first 140 moments). Solid
line: NLO distribution; dashed line: LO distribution; dotted line: expansion
to order $\alpha _{S}^{2}$ of the exponent in the NLO distribution.

\noindent Equation\thinspace (\ref{bella}) is our main result. Let us
comment on it. The terms most affected by long-distance effects are those
containing the combination $1-2w$ and they are the same in QCD and in the
ET. As a consequence, they cancel in the difference, proving that the
coefficient function \ is short-distance-dominated to NLO accuracy; we
believe that this cancellation occurs to any order in PT. The coefficient
function is meaningful up to 
\begin{equation}
n\,\ll \,n_{crit}^{\prime }\,=\,\frac{\mu }{Q}\,\exp \left[ \frac{1}{\beta
_{0}\,\alpha _{S}\left( Q^{2}\right) }\right] \,\sim \,\frac{\mu \,Q}{%
\Lambda ^{2}}.
\end{equation}
Note that $n_{crit}^{\prime }$ lies between the critical values for the soft
terms and the hard collinear one: 
\begin{equation}
n_{crit}\ll n_{crit}^{\prime }\ll n_{crit}^{2}.
\end{equation}
The coefficient function contains also the jet factor, entirely coming from
the QCD form factor: 
\begin{equation}
J_{n}=\exp \left[ \frac{C_{1}}{\beta _{0}}\log (1-w)\right] .
\end{equation}

\psfig{bbllx=100pt, bblly=310pt, bburx=600pt, bbury=740pt,
file=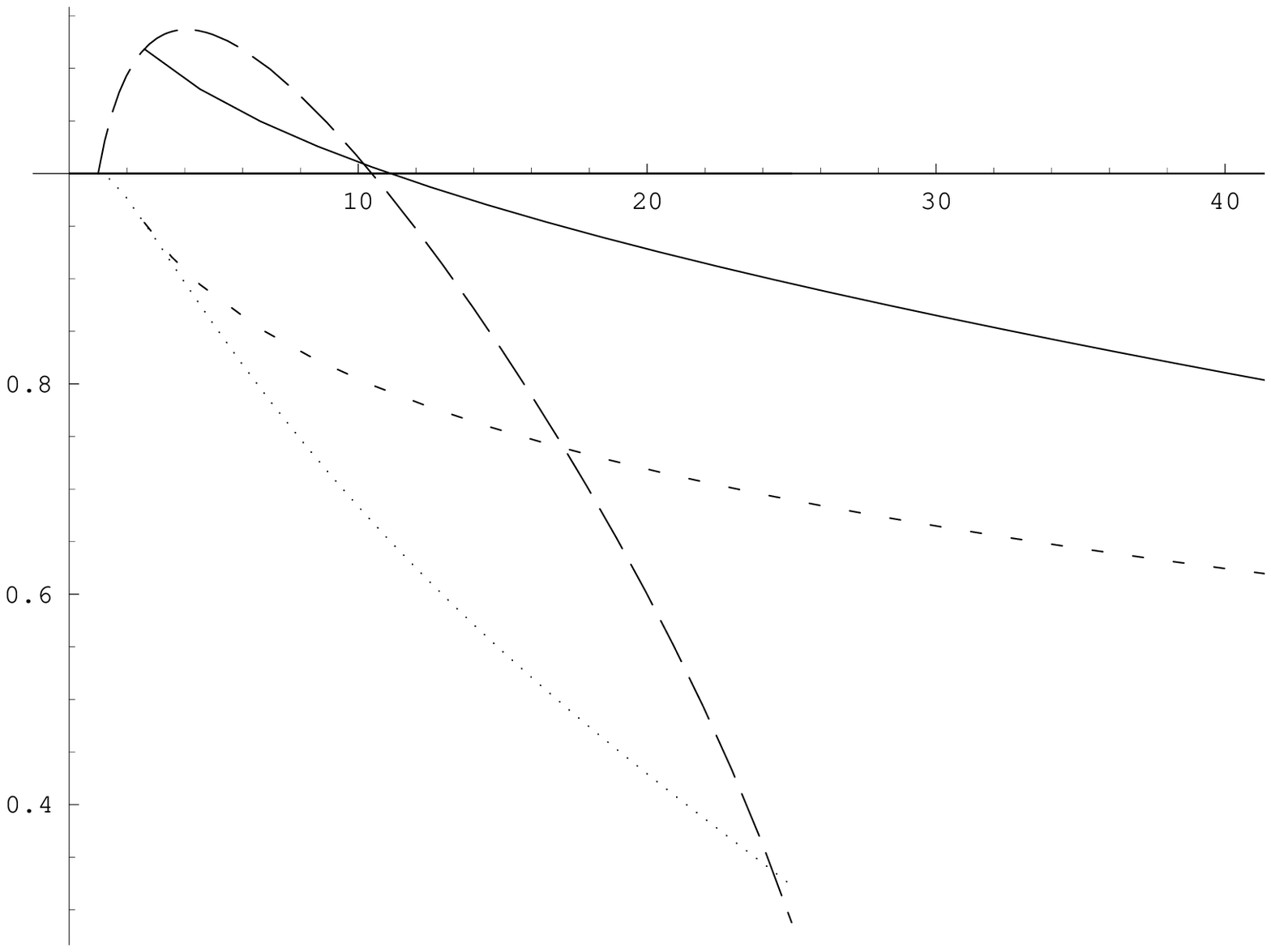, height=8cm, width=13cm} \vspace{-1cm}

\noindent Figure~5: Comparison of the coefficient function with the QCD form
factor. Solid line: NLO coefficient function; big dashed line: NLO QCD
distribution; small dashed line: LO coefficient function; dotted line: LO
QCD distribution.

\noindent The coefficient function is plotted in fig.\thinspace 4 for the
values of the parameters specified previously. It becomes singular at $%
n_{crit}^{\prime }\cong 297,$ i.e. an order of magnitude above the QCD form
factor or the shape function. For $n\lesssim 50,$ the NLO curve has a shape
similar to the LO one and it is shifted up with respect to the latter by $%
\sim 0.2.$ In fig.\thinspace 5 we compare the coefficient function with the
QCD form factor. Note that the large-$N$ approximation produces a cusp
singularity in the NLO coefficient function at $n=r.$

The computation of the shape function from first principles --- namely
lattice QCD --- requires a large amount of theoretical work, consisting of
the following steps:

\begin{enumerate}
\item  analytic continuation from Euclidean to Minkowski space of the
relevant correlation function, a 4-point function;

\item  lattice regularization of the effective theory in which the shape
function is defined, which involves the HQET (Wilson lines \textit{off} the
light cone) and the LEET (Wilson lines \textit{on} the light-cone);

\item  computation of the shape function in lattice perturbation theory to
full order $\alpha _{S};$

\item  matching of the lattice shape function with the QCD distribution to
full order $\alpha _{S};$

\item  Numerical computation with a Monte Carlo program of the correlation
function on the lattice.
\end{enumerate}

\noindent The first step was taken in ref.\thinspace \cite{lattice}, showing
the possibility of a lattice computation of the shape function. Step 2 has
only partially been completed: the lattice HQET has been formulated in ref. 
\cite{ancoraio} while the lattice LEET --- as far as we know --- has not
been constructed yet. The main difference is that, in the latter case, not
only soft singularities but also light-cone singularities appear in the
euclidean correlation functions; we do not attempt at the construction of
the lattice LEET here. No theoretical work exists --- as far as we know ---
on points 3, 4 and 5. In general, a lot of work is yet to be done: the
purpose of this note was simply showing how to resum to all orders the
threshold logarithms appearing in the coefficient function.

Our result (\ref{bella}) applies to the coefficient function of the shape
function defined in lattice regularization\thinspace ,\ after the
identification is done: 
\begin{equation}
\frac{1}{a}=c\,\mu ,
\end{equation}
where $c$ is a constant or order 1 and $a$ is the lattice spacing. $c$ is
determined, for example, imposing the equality of the coefficients of the
single logarithms in the lattice shape function and in eq.\thinspace (\ref
{cheneso}) (steps 3 and 4). In our work, we have \textit{assumed} the
consistency of the lattice LEET and the \textit{knowledge} of the one-loop
lattice amplitude.

In the absence of a (non-perturbative) computation of the shape function, we
can compare our coefficient function evaluated with a cut-off of a few times
the hadronic scale (in practice $1\div 2$ GeV) directly with the
experimental data. The mismatch is the non-perturbative component up to a
scale $\mu ,$ namely the shape function. The factorization scale $\mu $ acts
effectively in the coefficient function as a prescription for the Landau
pole. In this respect, our coefficient function can be considered as a
complete QCD form factor, with a prescription to remove the Landau pole
outside region (\ref{cinshape}). As we have seen, the singularity is not
completely removed, but it is shifted to much larger values of $N.$ We
stress that our splitting of the QCD form factor into a coefficient function
and a shape function represents a consistent separation of perturbative from
non-perturbative effects; it somehow differs from the usual strategy of
finding a good prescription for the running coupling in the non-perturbative
region.

\section{Conclusions}

We have presented a NLO evaluation of the resummed coefficient function for
the shape function. Our result, together with a one-loop lattice computation
that is still missing, allows relating the shape function computed with
lattice QCD to distributions in heavy-flavour decays in the semi-inclusive
region (\ref{cinshape}). The NLO analysis corroborates the results of the
leading order one, but it does not reveal any new qualitative feature. The
coefficient function is short-distance-dominated in the relevant region (\ref
{cinshape}), and corrections to factorization are expected to be of the
order $\Lambda /\mu ,$ i.e. to involve one inverse power of the
factorization scale. The NLO coefficient function contains also a jet factor
that takes into account the emission of hard collinear gluons. The latter
process is indeed not described by the effective theory, which takes into
account only soft emission up to the scale $\mu .$ On the quantitative side,
NLO effects increase the coefficient function by $\sim 20$\% for $\mu =2$
GeV and $n\lesssim 50$ in the case of $B$ decays.

In general, the process (\ref{generalizzo}) contains three different scales:%
\footnote{%
I wish to thank G. Korchemsky for a discussion on this point.} 
\begin{eqnarray}
\qquad \qquad \qquad \qquad \qquad \qquad \qquad \qquad i)\,\,\,\quad
\,Q^{2}\, &\gg &\,\Lambda ^{2},  \notag \\
ii)\,\,\,\,Q\,k_{+}\, &\sim &\,Q\,\Lambda \,\gg \,\Lambda ^{2},  \notag \\
iii)\,\,\,\,\,\,\,\,k_{+}{}^{2}\, &\sim &\,\Lambda ^{2}.
\end{eqnarray}
The emission of gluons with transverse momenta ranging between $i)$ and $ii)$
is reliably computed in perturbation theory, while fluctuations in region $%
iii)$ ask for a non-perturbative treatment. With factorization by means of
the shape function, another scale is introduced in the problem, intermediate
between $ii)$ and $iii):$%
\begin{equation}
iv)\,\,\,\,\mu \,k_{+}\sim \mu \,\Lambda \,\gg \,\Lambda ^{2}.\quad
\end{equation}
The fluctuations between scales $iii)$ and $iv)\,$\ are factorized in the
shape function. The general idea behind the shape-function approach is that
soft contributions are more ``non-perturbative'' than collinear ones in
double-logarithmic problems in region (\ref{cinshape}), because they involve
smaller transverse momenta. These soft non-perturbative effects can be
represented as hadronic matrix elements of non-local operators involving
Wilson lines. For a given process, one identifies a dominant kinematical
configuration and replaces the hard partons with Wilson lines with the same
momenta.

To summarize, we have presented a NLO evaluation of the resummed coefficient
function of the shape function, which confirms the consistency of the
effective theory approach and allows a more accurate theoretical analysis.

\begin{center}
Acknowledgements
\end{center}

I wish to thank in particular G. Ricciardi for having checked part of the
computation and S. Catani for suggestions. I also thank A. Banfi, M.
Cacciari, M. Ciafaloni, M. Ciuchini, D. De Florian, P. Gambino, E.\thinspace
Gardi, M. Grazzini, T. Hurth and P. Nason for discussions.

\end{document}